\documentclass[pdflatex,sn-mathphys]{sn-jnl}% Math and Physical Sciences Reference Style
\jyear{2021}%

\theoremstyle{thmstyleone}%
\theoremstyle{thmstyletwo}%

\theoremstyle{thmstylethree}%

\newcommand{\nbse}   {${\rm NbSe_2}$}
\newcommand{\nii}   {${\rm NiI_2}$}

\begin{document}

\title[Article Title]{Field-Resilient Supercurrent Diode in a Multiferroic Josephson Junction}

%%=============================================================%%
%% Prefix	-> \pfx{Dr}
%% GivenName	-> \fnm{Joergen W.}
%% Particle	-> \spfx{van der} -> surname prefix
%% FamilyName	-> \sur{Ploeg}
%% Suffix	-> \sfx{IV}
%% NatureName	-> \tanm{Poet Laureate} -> Title after name
%% Degrees	-> \dgr{MSc, PhD}
%% \author*[1,2]{\pfx{Dr} \fnm{Joergen W.} \spfx{van der} \sur{Ploeg} \sfx{IV} \tanm{Poet Laureate} 
%%                 \dgr{MSc, PhD}}\email{iauthor@gmail.com}
%%=============================================================%%

\author*[1]{\fnm{Hung-Yu} \sur{Yang}}\email{hungyuyang@ucla.edu}

\author[2]{\fnm{Joseph J.} \sur{Cuozzo}}

\author[1,3]{\fnm{Anand Johnson} \sur{Bokka}}
\author[1]{\fnm{Gang} \sur{Qiu}}
\author[1]{\fnm{Christopher} \sur{Eckberg}}
\author[4]{\fnm{Yanfeng} \sur{Lyu}}
\author[5]{\fnm{Shuyuan} \sur{Huyan}}
\author[5,6]{\fnm{Ching-Wu} \sur{Chu}}
\author[7]{\fnm{Kenji} \sur{Watanabe}}
\author[8]{\fnm{Takashi} \sur{Taniguchi}}
\author*[1]{\fnm{Kang L.} \sur{Wang}}\email{wang@ee.ucla.edu}

\affil[1]{\orgdiv{Department of Electrical and Computer Engineering}, \orgname{University of California, Los
Angeles}, \state{CA}, \country{USA}}

\affil[2]{\orgdiv{Materials Physics Department}, \orgname{Sandia National Laboratories}, \orgaddress{\city{Livermore}, \state{CA}, \country{USA}}}

\affil[3]{\orgdiv{Department of Materials Science and Engineering}, \orgname{University of California, Los
Angeles}, \state{CA}, \country{USA}}

\affil[4]{\orgdiv{School of Science}, \orgname{Nanjing University of Posts and Telecommunications}, \state{Nanjing}, \country{China}}

\affil[5]{\orgdiv{Department of Physics and Texas Center for Superconductivity}, \orgname{University of Houston, Houston}, \state{TX}, \country{United States}}

\affil[6]{\orgname{Lawrence Berkeley National Laboratory, Berkeley}, \state{CA}, \country{United States}}

\affil[7]{\orgname{Research Center for Electronic and Optical Materials, National Institute for Materials Science}, \state{Tsukuba}, \country{Japan}}
% Research Center for Electronic and Optical Materials, National Institute for Materials Science, 1-1 Namiki, Tsukuba 305-0044, Japan
\affil[8]{\orgname{Research Center for Materials Nanoarchitectonics, National Institute for Materials Science}, \state{Tsukuba}, \country{Japan}}
% Research Center for Materials Nanoarchitectonics, National Institute for Materials Science,  1-1 Namiki, Tsukuba 305-0044, Japan

%%==================================%%
%% sample for unstructured abstract %%
%%==================================%%

\abstract{
The research on supercurrent diodes has surged rapidly due to their potential applications in electronic circuits at cryogenic temperatures. To unlock this functionality, it is essential to find supercurrent diodes that can work consistently at zero magnetic field and under ubiquitous stray fields generated in electronic circuits. However, a supercurrent diode with robust field tolerance is currently lacking.  
Here, we demonstrate a field-resilient supercurrent diode by incorporating a multiferroic material into a Josephson junction. We first observed a pronounced supercurrent diode effect at zero magnetic field. More importantly, the supercurrent rectification persists over a wide and bipolar magnetic field range beyond industrial standards for field tolerance. By theoretically modeling a multiferroic Josephson junction, we unveil that the interplay between spin-orbit coupling and multiferroicity underlies the unusual field resilience of the observed diode effect.
This work introduces multiferroic Josephson junctions as a new field-resilient superconducting device for cryogenic electronics.
}

\maketitle

% \section*{Main}\label{sec_intro}

Semiconductor diodes are fundamental electronic components crucial for rectifying, regulating, and controlling the flow of electrical current in electronic circuits and systems, playing a pivotal role in the functionality of a wide range of devices from power supplies to digital electronics \cite{sze_physics_2021}. Supercurrent diodes, which rectify the zero-resistance supercurrent in superconductors, play key functions in digital electronics at cryogenic temperatures. For example, in an electronic flip-flop memory, a binary bit can be represented by the current going through one arm or the other; this can be achieved similarly by placing supercurrent diodes on each arm and controlling their rectification directions \cite{matisoo_tunneling_1967}. More importantly, for a cryogenic memory application, the readout can be done through the supercurrent diode effect (SDE) that in principle leads to low power consumption and an infinite on/off ratio, thanks to the zero resistance in the superconducting state \cite{buck_cryotronsuperconductive_1956,matisoo_tunneling_1967,alam_cryogenic_2023}.

In the past few years, supercurrent diodes have been found extensively in various systems under a magnetic field~\cite{ando_observation_2020, baumgartner_supercurrent_2021, Baumgartner_2022, Bauriedl2022, pal_josephson_2022, hou_ubiquitous_2023, Ciaccia2023, Costa2023,nadeem_superconducting_2023} while only few work at zero magnetic field. Among the zero-field supercurrent diodes \cite{qiu_emergent_2023,diez-merida_symmetry-broken_2023,lin_zero-field_2022,zhao_time-reversal_2023,wu_field-free_2022,yu_time_2024,narita_field-free_2022,jeon_zero-field_2022}, most of them require a magnetic field to polarize the ferromagnetic component and initialize the diode; the ferromagnetism grants the field-tunability to these diodes, while also makes them unable to work persistently over bipolar magnetic fields.
For practical applications, ubiquitous stray fields in a common circuit environment (up to 10 mT) can easily flip the supercurrent rectification direction and make this type of diodes unreliable \cite{khan_novel_2018}. Currently, a clear strategy for field-resilient supercurrent diodes that can work at zero magnetic field and tolerate stray fields in electrical circuits remains lacking.

The SDE is governed by the symmetry properties; the breaking of inversion and time-reversal symmetries simultaneously is essential for SDE regardless of the material platform \cite{yuan_supercurrent_2022,davydova_universal_2022,daido_intrinsic_2022}. For example, a 2D superconductor with Rashba spin-orbit coupling (RSOC) breaking the inversion symmetry, and an applied in-plane transverse magnetic field breaking time-reversal symmetry, exhibits SDE \cite{yuan_supercurrent_2022}. In this study, we employed NiI$_2$, a 2D multiferroic material into a van der Waals (vdW) Josephson junction (JJ) to create a field-resilient supercurrent diode. The coexisting spiral magnetic order and ferroelectric order in NiI$_2$ naturally break both inversion and time-reversal symmetry (Fig. \ref{fig:fig1}a) \cite{tokura_multiferroics_2014,billerey_neutron_1977,kuindersma_magnetic_1981,kurumaji_magnetoelectric_2013}, presumably satisfying symmetry requirements for SDE. Furthermore, the coupling between magnetic and electric orders makes a multiferroic more robust against the magnetic field (e.g. coercivity enhancement) \cite{martin_nanoscale_2008,heron_electric-field-induced_2011}, granting the field-resilience for SDE. Lastly, the strong magnetoelectric coupling in multiferroics enables controllable switching of magnetic order \cite{vaz_electric_2012,heron_deterministic_2014,meisenheimer_switching_2024} and potentially the switching of SDE by electrical gates. Incorporating this non-volatility and gate tunability into supercurrent diodes could open the door to practical cryogenic memory devices.

\section*{Zero-field SDE in a multiferroic vdW JJ}\label{sec_zero_field_SDE}

\begin{figure}[t]
    \includegraphics[width=\columnwidth]{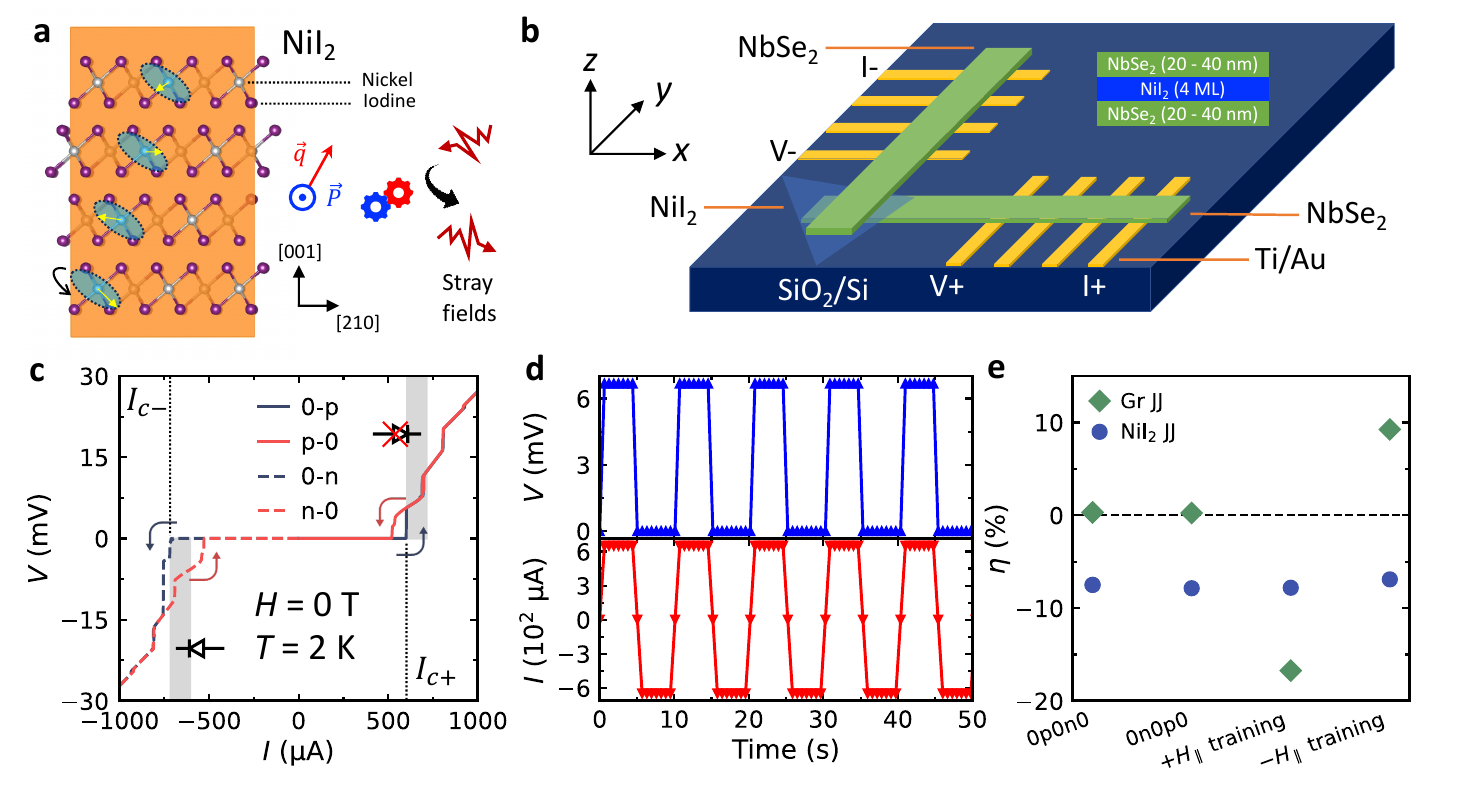}
    \centering
    \caption{\textbf{2D Multiferroic NiI$_2$ and zero-field supercurrent diode effect in the NiI$_2$ vdW JJ.} \textbf{a}, Crystal structure and multiferroic order of NiI$_2$, which consists of a spiral magnetic order (described by the wave vector $\vec{\mathbf{q}}$ \cite{kuindersma_magnetic_1981}) and an in-plane ferroelectric order ($\vec{\mathbf{P}}$ \cite{kurumaji_magnetoelectric_2013}). The yellow arrow on the Ni atoms represents the spin direction and the shaded area represents the spin spiral plane. \textbf{b}, Device geometry of the NiI$_2$ JJ. \textbf{c}, $V-I$ characteristic of the NiI$_2$ JJ. 0-p, p-0, 0-n, and n-0 refer to curves with current sweeping from 0 $\mu$A to $+1000$ $\mu$A, $+1000$ $\mu$A to 0 $\mu$A, 0 $\mu$A to $-1000$ $\mu$A, and $-1000$ $\mu$A to 0 $\mu$A. The critical current $I_{c+}$ (600 $\mu$A) and $I_{c-}$ (718 $\mu$A) are defined by the first critical jump in $V$ in the 0-p and 0-n (switching) curves, respectively. \textbf{d}, Demonstration of supercurrent rectification with $I_\text{bias}=\pm 650$ $\mu$A. \textbf{e}, Comparison of zero-field supercurrent diode rectification efficiency ($\eta=\frac{I_{c+}-\lvert I_{c-}\rvert}{I_{c+}+\lvert I_{c-}\rvert}$) between the Gr JJ and the NiI$_2$ JJ under different current-sweeping and field-training protocols. The magnetic field was set to oscillate to zero from 3 T before performing the 0p0n0 and 0n0p0 measurements. The measurements for 0p0n0 and 0n0p0 tests are repeated five times to acquire error bars, which are smaller than the marker size for both cases. The 0p0n0 and 0n0p0 refer to opposite current-sweeping protocols, where a positive bias current is applied first in the 0p0n0 measurement and a negative bias current is applied first in the 0n0p0 measurement, respectively.
    The training fields $\pm H_\parallel = \pm 1$ T were used for both devices. The field training was performed at $T=10$ K.}  
    \label{fig:fig1}
\end{figure}

Since the multiferroic order in NiI$_2$ persists down to the 2D monolayer (ML) limit \cite{song_evidence_2022,fumega_microscopic_2022,amini_atomic-scale_2023}, we exfoliated a NiI$_2$ flake of 4 MLs thick to facilitate the Josephson coupling while keeping the multiferroic order. It is then re-assembled with two NbSe$_2$ flakes to make a NbSe$_2$/NiI$_2$/NbSe$_2$ vertical vdW JJ (NiI$_2$ JJ in short), thanks to the freedom to manipulate vdW materials with the 2D transfer assembly technique (Fig. \ref{fig:fig1}b, see also Methods and Extended Data Fig. \ref{fig:extd_device}). 
Figure \ref{fig:fig1}c shows a typical $V-I$ characteristic of the NiI$_2$ JJ. The quantities relevant to SDE are the critical current for opposite bias directions, $I_{c+}$ and $I_{c-}$, at which the JJ transitions from a superconducting state to a normal state.
The critical current difference $\Delta I_c \equiv I_{c+}- \lvert I_{c-} \rvert = -118$ $\mu$A and diode rectification efficiency $\eta \equiv \frac{I_{c+}- \lvert I_{c-} \rvert}{I_{c+}+ \lvert I_{c-} \rvert} \sim -8 \%$ were obtained. The difference in magnitude also allows us to define a diode working range (gray stripes in Fig. \ref{fig:fig1}c), within which the supercurrent only flows in one direction but not the other (Fig. \ref{fig:fig1}d). 
The consistent switching with repetitive current biasing cycles (see also Extended Data Fig. 2) shows the robustness of the SDE in the NiI$_2$ JJ at zero field.

To further confirm the zero-field SDE in the multiferroic NiI$_2$ JJ, a NbSe$_2$/few-layer graphene/NbSe$_2$ vdW JJ (Gr JJ) was fabricated as a reference device, and different tests were performed to confirm the observed zero-field SDE is intrinsic 
(Fig. \ref{fig:fig1}e). If the heating effect is significant, the rectification efficiency should flip sign between these two measurements \cite{wu_field-free_2022}. We found that the diode rectification efficiency $\eta$ of the Gr JJ stayed near zero and that of NiI$_2$ JJ remained $\sim -8 \%$, showing the heating effect was insignificant in the SDE of both devices.

Next, opposite in-plane fields ($\pm H_\parallel$) were applied to train the magnet and devices, and then $V-I$ characteristics were measured at zero field. Here, we utilized the following fact to simulate the effect of stray fields: With a large positive (negative) magnetic field up to several tesla being applied through the magnet of our measurement system, a small negative (positive) remnant field on the order of $\sim$ 1 mT could remain after the field is set to zero \cite{qdappnote_2009}. For a field-resilient supercurrent diode, $\eta$ must not flip its sign for opposite training fields to continue rectifying supercurrent in the same direction under stray fields. As shown in Fig. \ref{fig:fig1}e, the $\eta$ of NiI$_2$ JJ surprisingly remained negative $\sim -8\%$ after both $\pm H_\parallel$ training, in strong contrast to the Gr JJ where its $\eta$ flipped the sign between the $\pm H_\parallel$ training. As will be discussed in Fig. \ref{fig:fig2}, our Gr JJ exhibits a pronounced SDE with anti-symmetric field dependence, with $\eta \sim \pm 20 \%$ for $H_\parallel \sim  \pm 1 $ mT, respectively. The nonzero but opposite $\eta$ values of the Gr JJ after $\pm H_\parallel$ training are thus false-positive zero-field SDE and are a result of remnant fields induced by the field training, in agreement with Fig. \ref{fig:fig2}. The tests and comparison demonstrated in Fig. \ref{fig:fig1}e show that the zero-field SDE in the multiferroic NiI$_2$ JJ is not only intrinsic but also field-resilient. In the NiI$_2$ JJ, $\eta$ also surpasses the values reported among the systems that do not require a field initialization for the zero-field SDE ($\eta_{\text{max}} \sim 3 \%$ in both Fe(Te,Se)/Fe(Te,Se) vdW JJ \cite{qiu_emergent_2023} and NbSe$_2$/Nb$_3$Br$_8$/NbSe$_2$ vdW JJ \cite{wu_field-free_2022}).

\section*{Field resilience of the SDE in NiI$_2$ JJ}\label{sec_field_resistant_SDE_IP}

\begin{figure}[t]
    \includegraphics[width=0.7\columnwidth]{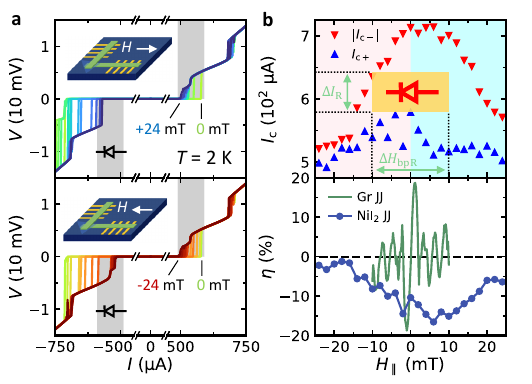}
    \centering
    \caption{\textbf{In-plane field dependence of the supercurrent diode effect in the NiI$_2$ JJ.} \textbf{a}, Top panel: $V-I$ characteristic of the NiI$_2$ JJ with 0 mT $<H_\parallel<$ 24 mT, with a 2 mT field increment. Bottom panel: $V-I$ characteristic of the NiI$_2$ JJ with -24 mT $<H_\parallel<$ 0 mT. All plotted curves are switching curves (0-p and 0-n sweeps). \textbf{b}, Top panel: critical current $I_{c+}$ and $\lvert I_{c-}\rvert$ as a function of $H_\parallel$. The pink and cyan background represents the negative and positive field range, respectively. The yellow block marks the bipolar working field range of the supercurrent diode between $\pm10$ mT with a figure of merit $F_\text{R} = \Delta I_\text{R} \times \Delta H_\text{bpR} \sim 10^3\ \text{mT}\cdot \mu \text{A}$. Bottom panel: $\eta$ as a function of $H_\parallel$ of Gr JJ and NiI$_2$ JJ.}
    \label{fig:fig2}
\end{figure}

To demonstrate the robustness of zero-field SDE in the NiI$_2$ JJ under stray fields, we measured the field dependence of the SDE. The $V-I$ characteristics of the NiI$_2$ JJ were acquired with in-plane fields ($H_\parallel$) ranging from $+24$ mT to 0 mT (Fig. \ref{fig:fig2}a, top panel), and 0 mT to $-24$ mT (bottom panel), over which the multiferroic order persists (Extended Data Fig. \ref{fig:extd_iets}). A consistent negative diode rectification efficiency was observed for all fields, with negative bias-induced critical transitions occurring beyond the gray stripe (Fig. \ref{fig:fig2}a), defined as the range of critical transitions for positive bias. The unidirectional supercurrent rectification over $H_\parallel = \pm 24$ mT is directly linked to an unusual symmetric field dependence of SDE. Such a field dependence defies the typical anti-symmetric field dependence of SDE, where an external magnetic field solely controls the time-reversal symmetry. \cite{ando_observation_2020,baumgartner_supercurrent_2021,jeon_zero-field_2022,hou_ubiquitous_2023}.

The symmetric in-plane field dependence is further shown by extracting $I_{c+}$ and $\lvert I_{c-} \rvert$ at each field (Fig. \ref{fig:fig2}b, top panel) and calculating their corresponding $\eta$ (bottom panel). Again, the data points representative of $\lvert I_{c-} \rvert$ are always above $I_{c+}$ between $H_\parallel=\pm 24$ mT, showing a robust negative SDE regardless of reversing the magnetic field direction. 
Importantly, $\eta$ consists of a symmetric, dome-shaped field dependence. Such a predominantly symmetric field dependence makes it possible to draw the widest bipolar diode working range reported so far over $\pm 10$ mT ($\sim$ 8000 A/m, the maximum field tolerance of industrial MRAM devices manufactured by Everspin \cite{everspin_mram_2017}), where we can use the same amount of current biased in the opposite directions to rectify the supercurrent. Thus, a bipolar figure of merit can be defined as $F_\text{R} \equiv \Delta I_\text{R}\ (\text{current rectification range}) \times \Delta H_\text{bpR}\ (\text{bipolar field rectification range})$, as the area of the yellow block shown in Fig. \ref{fig:fig2}b.  For our device, $F_\text{R}$ over $\pm 10$ mT is on the order of $10^3$ mT$\cdot$$\mu$A, which is two orders of magnitude larger than the existing supercurrent diode where a bipolar diode working range may be defined (the maximum of $F_\text{R}$ in NbSe$_2$/Nb$_3$Br$_8$/NbSe$_2$ JJ is about $10^1$ mT$\cdot$$\mu$A \cite{wu_field-free_2022}).

We highlight the unique field-resilient SDE in the NiI$_2$ JJ by comparing its field dependence to the Gr JJ (Fig. \ref{fig:fig2}b, bottom panel). 
For the Gr JJ, the SDE therein exhibits an anti-symmetric field dependence of $\eta$ with multiple sign changes, corresponding to the lobes of the Fraunhofer pattern developing as the field increases (see Extended Data Fig. \ref{fig:extd_Fpattern}a). The maxima of $I_c$ for opposite current biases shift from zero to opposite fields due to the self-field effect induced by the cross-junction geometry, which further leads to an SDE \cite{ferrell_self-field_1963,yamashita_magneticfield_1967}.
We expect the field-anti-symmetric SDE \emph{without} a bipolar working range to be typical of vdW JJ with a nonmagnetic barrier and a cross-junction geometry. On the contrary, the interference pattern of the NiI$_2$ JJ was ``truncated" for the positive current bias, while preserved for the negative current bias (see Extended Data Fig. \ref{fig:extd_Fpattern}b). The pattern thus leads to a persistent negative $\eta$, with a symmetric field dependence and a wide bipolar diode working range establishing the field-resilient SDE in NiI$_2$ JJ. We have also examined the SDE under out-of-plane magnetic fields and observed again a persistent negative $\eta$ in NiI$_2$ JJ with reduced efficiency, contrary to the reference device showing an anti-symmetric field dependence of SDE (Extended Data Fig. \ref{fig:extd_oopH}).

\section*{Non-monotonic temperature dependence of SDE}\label{sec_SDE_temp_dep}

\begin{figure}[t]
    \includegraphics[width=0.7\columnwidth]{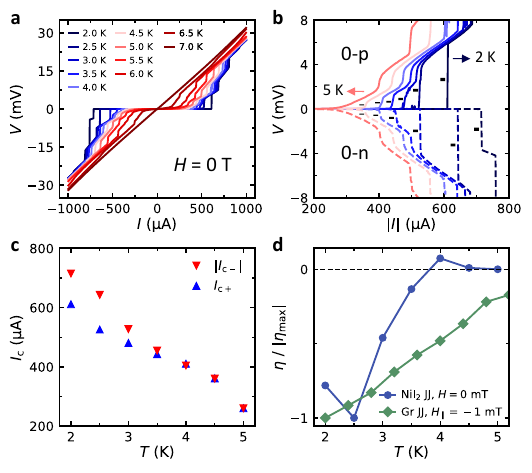}
    \centering
    \caption{\textbf{Non-monotonic temperature dependence of supercurrent diode effect in the NiI$_2$ JJ.} \textbf{a}, $V-I$ characteristic (switching curves) measured at different temperatures. \textbf{b}, $V-\lvert I \rvert$ characteristic recorded at $T\leq 5$ K. The solid and dashed lines represent 0-p sweep and 0-n sweeps, respectively. The critical transitions are pointed out by short black line segments with varied widths. \textbf{c}, $I_{c+}$ and $\lvert I_{c-}\rvert$ as a function of temperature. \textbf{d}, $\eta$ normalized by the maximum $\lvert \eta_{\text{max}} \rvert$ as a function of temperature for both NiI$_2$ JJ at zero field and Gr JJ at nonzero field. $\eta_{\text{max}}$ is -10 \% and -20 \% for NiI$_2$ and Gr JJ, respectively.}
    \label{fig:fig4}
\end{figure}

Finally, we investigate the temperature dependence of SDE in the NiI$_2$ JJ. The results reveal its non-monotonic temperature dependence and a
sign change. The zero-field SDE at different temperatures from the $V-I$ characteristics is illustrated in Fig. \ref{fig:fig4}a. In Fig. \ref{fig:fig4}b, the 0-p and 0-n sweeps are compared to show their critical transitions for $T \leq 5$ K. The critical transition defining $I_c$ at each temperature is labeled by short black lines, where the same transition can be tracked up to $T=5$ K, the transition temperature of the JJ (see also Extended Data Fig. \ref{fig:extd_RT}).

Figure~\ref{fig:fig4}c presents the temperature dependence of $I_{c \pm}$ of the NiI$_2$ JJ at zero field, from which $\eta$ and $\lvert \eta_\text{max} \rvert$ are extracted and compared to that of the Gr JJ measured at $H_\parallel=-1$ mT in Fig. \ref{fig:fig4}d. For the Gr JJ, the SDE follows a monotonic temperature dependence where $\eta_\text{max}$ happens at the lowest temperature reached, similar to other non-multiferroic lateral JJs \cite{pal_josephson_2022,jeon_zero-field_2022}. However, in the NiI$_2$ JJ, two unusual things appear. First of all, $\eta_\text{max}$ appears at $T=2.5$ K, instead of $T=2$ K which is the lowest temperature reached. Secondly, after the enhancement of SDE at $T=2.5$ K, $\eta$ drops more quickly than expected and undergoes a sign change before it completely vanishes.
Below, we develop a theoretical model to capture our findings of SDE in the NiI$_2$ JJ, including its appearance at zero field, enhanced bipolar field resilience, and uncommon non-monotonic temperature dependence.

\section*{Theoretical Modeling}\label{sec_discuss}

%=============================================
%                 FIGURE
%=============================================
\begin{figure}[h!!!!]
    \centering
    \includegraphics[width=0.95\linewidth]{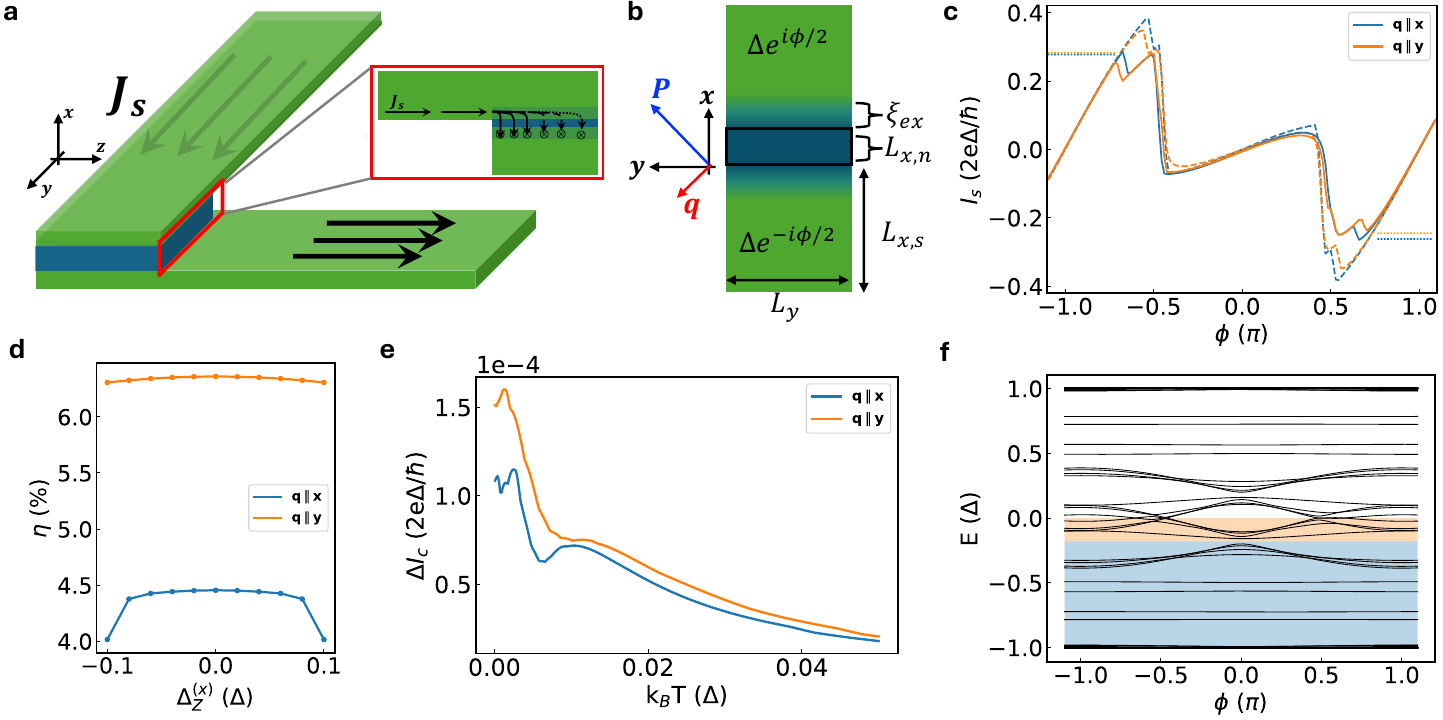} 
    \caption{\label{fig:theory}
    \textbf{Multiferroic JJ simulation.}
    \textbf{a}, Schematic of the cross junction device where the supercurrent density $\bf J_s$ tends to reside near the surfaces of the superconducting electrodes. 
    \textbf{b}, Schematic of the planar junction corresponding to the SC/helimagnet/SC cross-section marked by the red rectangle in panel $\bf a$. 
    \textbf{c}, The simulated junction CPR with (solid) and without (dashed) RSOC for $\bf q \parallel \bf x$ and $\bf q \parallel \bf y$. Unless otherwise stated, parameters used in simulations are: $\Delta = 0.4 t$, $\mu = 1.57 t$, $\alpha_R = 0.004ta$, $J_{exc} = 0.3t$, $U_{barrier}= 4t$, $\vert \mathbf{q} \vert = 0.01 \frac{\pi}{a}$, $L_{x,s} = 300a$, $L_{x,n} = 3a$, $L_y = 10a$, and $\xi_{exc} = 5a$ where $t = \frac{\hbar^2}{2m^* a^2}$ and $a$ is the tight binding lattice constant.
    \textbf{d}, Diode rectification efficiency $\eta$ versus Zeeman splitting along $\bf x$ with RSOC for $\bf q \parallel \bf x$ and $\bf q \parallel \bf y$.
    \textbf{e}, Critical current difference $\Delta I_c = I_{c +} - \vert I_{c -} \vert$ versus temperature with RSOC.
    \textbf{f}, The simulated Andreev bound state spectrum for $\bf q \parallel \bf y$ and $B = 0$.
    }
\end{figure}

In our \nii~JJ, electrons in \nbse~can experience spin-orbit interactions that are intrinsic to \nbse~and NiI$_2$ or arise from interfacial effects \cite{fumega_microscopic_2022,husain_emergence_2020}.
The geometry of the cross junction modifies the supercurrent density to reside near the surfaces of the two crossed \nbse~flakes (Fig.~\ref{fig:theory}a) \cite{stuehm_diffraction_1974}, which will enhance the role of RSOC in the Josephson coupling between the \nbse~flakes. For generality, we
focus on the role of RSOC in a generic cross JJ with a helimagnet weak link \cite{Martin2012, Hals2017, Hess2023}.
For a propagation vector 
$
{\bf q} = (q_x, q_y,0),
$
with helimagnet spin texture in real space given by ${\bf M} = M(-\sin({\bf q \cdot r}), \cos({\bf q \cdot r}), 0)$, we can write the Bogoliubov de-Gennes Hamiltonian in the momentum space as
\begin{align}
    H_{BdG} & = \frac{1}{2} \sum_{\bf k} \psi^{\dagger}_{\bf k}
    \begin{pmatrix}
        h(\bf k) - \mu & \Delta_{sc} \\
        \Delta_{sc}^* & \mu - T^{-1} h({\bf k}) T
    \end{pmatrix}
    \label{eq:bdgham}
    \psi_{\bf k} \\
    h(\bf k) & = \frac{\hbar^2 (\mathbf{k}^2+\mathbf{q}^2/4)}{2m^*}  + \frac{\hbar^2 (\bf q \cdot \bf k)}{2m^*} \sigma_z + J_{exc} \sigma_y + \alpha_R \left(k_y \sigma_x - k_x \sigma_y \right),
    \label{eq:ham}
\end{align}
where 
$
\psi_{\bf k}=(c_{\bf k \uparrow},c_{\bf k \downarrow},-c^{\dagger}_{-\bf k \downarrow},c^{\dagger}_{-\bf k \uparrow})^T
$
is a spinor of electron creation (annihilation) operators $c^{\dagger}_{\bf k \sigma}$ ($c_{\bf k \sigma}$) with momentum $\bf k$ and spin $\sigma$, $\Delta_{sc} = \Delta e^{i \phi} \sigma_x$ is the superconducting gap with phase $\phi$, $\mu$ is the chemical potential,
$\hbar$ is Planck's constant divided by $2\pi$, $m^*$ is the effective electron mass, $\alpha_R$ is the RSOC strength, and $J_{exc}$ is the exchange interaction energy. Here $\sigma_i$ are Pauli matrices and $T=i\sigma_y K$ is the time-reversal operator with complex conjugation operator $K$. 
In Eq.~(\ref{eq:ham}), the exchange spin splitting $J_{exc} \sigma_y$ arises from $M>0$ breaking TRS, and the $({\bf q \cdot k})$ term is associated with the spin-orbit coupling induced by the spin texture ${\bf M}({\bf r})$.
We discretize the Hamiltonian in Eq.~(\ref{eq:bdgham}) and perform numerical simulations of a helimagnetic JJ shown in Fig.~\ref{fig:theory}b. Using tight-binding simulations, we calculate the Andreev bound state spectrum of the JJ to find its current-phase relationship (CPR). 
To model the \nii~tunnel barrier, we include a potential barrier 
$
h_{b} = U_{barrier} \delta(x).
$
Additional details are described in the Methods section.

In Fig.~\ref{fig:theory}c, we present the CPR with and without RSOC for $\bf q$ oriented along $\bf x$ and $\bf y$. The global extrema of the CPR correspond to $I_{c \pm}$. We see $I_{c+} = \vert I_{c-} \vert$ when $\alpha_R = 0$. 
The absence of SDE here is due to the exchange interaction generating an effective Zeeman field which is perpendicular to the spin-orbit interaction originating from the spiral spin texture. Since the broken TRS and inversion symmetries are along orthogonal axes, a non-reciprocal supercurrent cannot develop~\cite{Hess2023}.
When $\alpha_R >0$, this no longer holds and an SDE develops with a maximum efficiency of $\sim 6\%$ when $\bf q \parallel \bf y$. Thus, the combination of helimagnetism and RSOC in the JJ is sufficient to result in a zero-field SDE. We have also performed a phenomenological depairing momentum analysis that leads to a similar conclusion (see Methods and Extended Data Fig. \ref{fig:depairing_q} for details).

Next, we consider the effects of an external magnetic field and discuss how a symmetric-in-field SDE generally emerges in a helimagnetic JJ.
To simulate the effects of a magnetic field, we consider an additional term in Eq.~(\ref{eq:ham}):
$
h_Z = g \mu_B (\mathbf{B} \cdot \sigma) = \Delta_Z/2 (\mathbf{\hat{B}} \cdot \sigma),
$ where $\bf \hat{B}$ is a unit vector parallel to $\bf B$.
When the helimagnet spin texture and RSOC coexist in the JJ, $\eta$ is an \textit{even} function of $\Delta_Z$ if $h_Z$ \textit{anti-commutes} with the terms in the Hamiltonian that are: (i) linear in $\bf k$  parallel to the current direction and (ii) proportional to $J_{exc}$.
When $h_Z$ obeys both anti-commutation relations, the BdG spectral gap closes symmetrically with $\pm \Delta_Z$ and $\eta$ is a purely even function of $h_Z$ (see Methods).
Indeed, as calculated in Fig.~\ref{fig:theory}d, a Zeeman splitting along the current direction (${\bf B} = B \bf x$) leads to a \textit{symmetric} modulation of $\eta$ in $\Delta_Z$, i.e. a symmetric-in-field SDE, regardless of the orientation of $\bf q$. For ${\bf B} = (0,B_y, B_z)$, $\eta$ will generally have a mixed functional dependence on the applied field (Extended Data Fig. \ref{fig:abs}). However, we emphasize that the symmetric field dependence is ubiquitous in helimagnetic JJs and is the key to the field resilience in NiI$_2$ JJ (see also Extended Data Fig. \ref{fig:depairing_q}). This is in strong contrast to other sources of non-reciprocal switching currents associated with magnetochiral anisotropy (MCA) \cite{Bauriedl2022, Baumgartner_2022, Costa2023}, finite-momentum superconductivity~\cite{He_2022, daido_intrinsic_2022, yuan_supercurrent_2022}, Meissner currents \cite{davydova_universal_2022, hou_ubiquitous_2023} or self-field effects~\cite{stuehm_diffraction_1974, Barone1975, yamashita_magneticfield_1967}, which are predicted to result in antisymmetric field dependence of $\eta$ as we demonstrated in the Gr JJ reference device.

The simulated temperature scaling of $\Delta I_c$ at zero field for helimagnetic JJs is presented in Fig.~\ref{fig:theory}e; the simulations reveal a non-monotonic behavior regardless of the direction of $\textbf{q}$. 
The exchange interaction associated with the helimagnet pushes the JJ close to a 0-$\pi$ transition where a significant second harmonic contribution develops (orange bands in Fig.~\ref{fig:theory}f) due to Andreev bound state energy level crossings at zero energy~\cite{Yokoyama2014}. Since these states lie near zero energy, their contribution to the CPR is more quickly washed out at finite temperatures compared to even lower energy states (blue bands in Fig.~\ref{fig:theory}f) which favor a $\phi = 0$ ground state. The competition between the supercurrent carried by states near zero energy and that by lower states leads to the non-monotonic scaling shown in Fig.~\ref{fig:theory}e. 
It is noted that the scaling behavior depends on the details of Andreev bound states and may be modified as other parameters change (see Extended Data Fig. \ref{fig:wideJJ})~\cite{Kokkeler2022, Lu2023}.

Lastly, we discuss the effects of an electric polarization $\bf P$ in a multiferroic JJ from our simulations (see Methods and Extended Data Fig. \ref{fig:ABSmultiJJ} for details). Between $\pm \bf P \parallel y$ we see a change in the CPR, indicating that the SDE can be tuned by flipping $\bf P$. The latter case occurs because flipping $\bf P$ simultaneously flips $\bf q$ along the current direction, whose effect on the SDE is also revealed in Fig. \ref{fig:theory}d. 
In addition, it is noted that the ferroelectric order in NiI$_2$ is closely linked to the strong SOC of iodine atoms, which could enhance the SDE in multiferroic JJs. Our simulation suggests that tuning electric polarization could uniquely manipulate and enhance SDE in multiferroic JJs.

\section*{Conclusion}\label{sec_conclusion}

Our work presents the first demonstration of a field-resilient supercurrent diode using a multiferroic NiI$_2$ vdW JJ. The key observation lies in the supercurrent diode that operates persistently not only at zero field but also under bipolar magnetic fields matching with industrial standards for field tolerance. This invention overcomes the significant limitation in conventional supercurrent diodes, which are driven by external magnetic fields and are susceptible to disruption by stray fields. 
Our simulations qualitatively capture the main observations of zero-field SDE, field-resilient SDE, and non-monotonic temperature dependence of the SDE in NiI$_2$ JJ.
Our theoretical modeling suggests that the combination of RSOC with helimagnetism plays a key role in the SDE in NiI$_2$ JJ, and these features may prevail in helimagnetic JJs. We point out the possibility of manipulating and enhancing the SDE by electrical gating in multiferroic JJs, which is an exciting tuning knob to explore in the future. The discovery may lead to the technology development of multiferroic supercurrent diodes with higher performance and tunability, opening up new possibilities for practical applications in cryogenic electronic circuits.

%%===========================================================================================%%
%% If you are submitting to one of the Nature Portfolio journals, using the eJP submission   %%
%% system, please include the references within the manuscript file itself. You may do this  %%
%% by copying the reference list from your .bbl file, paste it into the main manuscript .tex %%
%% file, and delete the associated \verb+\bibliography+ commands.                            %%
%%===========================================================================================%%

\bibliography{references}% common bib file
%% if required, the content of .bbl file can be included here once bbl is generated
%%\input sn-article.bbl

%% Default %%
%%\input sn-sample-bib.tex%

\section*{Methods}
\subsection*{Crystal growth}

Single crystals of NiI$_2$ were grown by chemical vapor transport technique. The starting materials were mixed in stoichiometric ratio ($\text{Ni}:\text{I}_2=1:1$, 500 mg in total) and sealed in 7-inch long silica tubes under vacuum. The tubes were placed in a single-zone tube furnace, with one end at the center. The temperature of the furnace was set to 580 $^\circ$C at 3 $^\circ$C/minute, dwelt for 60 hours, and then set to room temperature at the same rate. Black single crystals formed at the cold end of the tubes as hexagonal thin flakes. X-ray diffraction patterns (Bruker D8 ECO) of the single crystals showed clear (003) characteristic peak at $2\theta =13.42$ degrees, in agreement with the crystal structure reported in Inorganic Crystal Structure Database (ICSD).

High-quality NbSe$_2$ single crystals were prepared using the iodine vapor transport method \cite{naito_electrical_1982}. Stoichiometric amounts of Nb (99.9\%; Alfa Aesar) and Se (99.5\%; Alfa Aesar) powders were sealed in an evacuated quartz tube (1/2 inch diameter) with 2 mg/cm$^3$ of iodine as the transport agent, and introduced horizontally into a tube furnace. The temperature was slowly increased to 725 $^\circ$C, maintained for 3 days, and followed by furnace cooling down to room temperature. The large platelet single crystal picked out from the resulting sample was sealed on one side in another quartz tube along with a new mixture of Nb, Se, and iodine on the other side and heated through the same procedure. Subsequently, large and high-quality NbSe$_2$ single crystals were obtained.

\subsection*{Device fabrication}

The bottom contact electrodes were fabricated on SiO$_2$/Si substrates using photolithography and electron beam evaporation (Ti/Au, 10/40 nm). Thin flakes of h-BN, NbSe$_2$, and NiI$_2$ were exfoliated onto SiO$_2$/Si substrates using Scotch tapes. To assemble the NiI$_2$ vdW JJ, a piece of h-BN was first picked by dry transfer technique \cite{kinoshita_dry_2019,qiu_emergent_2023}, using polypropylene carbonate (PPC) polymer spin-coated on polydimethylsiloxane (PDMS) as the stamp. Once the h-BN is picked up, other flakes of the JJ were picked up in the following order: top NbSe$_2$, NiI$_2$, and then bottom NbSe$_2$. Occasionally, the structure that the h-BN has picked up may be released on the next target flake and be heated to increase the cohesion between different layers to facilitate the pick-up process. After the entire stack of flakes was completed, the h-BN/NbSe$_2$/NiI$_2$/NbSe$_2$ vdW JJ structure was released on the bottom contacts and was ready for transport measurements without further fabrication. During the dry transfer process, the JJ area was covered by h-BN the entire time to prevent contamination due to polymer residue. The graphite/NiI$_2$/graphite tunnel junction was fabricated in the same way. For the Gr JJ and NbSe$_2$/NbSe$_2$ devices, the bottom NbSe$_2$ flakes were exfoliated onto SiO$_2$/Si substrates using Scotch tapes. Other flakes were exfoliated on PDMS and then transferred on top of the bottom NbSe$_2$ flake one by one. After the entire device stack was completed, it was picked up using PPC and released on the bottom contacts. The transfer process was performed in an Ar-filled glove box with H$_2$O and O$_2$ levels below 1 ppm using a home-built transfer stage. The two-point contact resistance between different pins was below 100 ohms.

\subsection*{Transport measurements}

Resistance and $V-I$ characteristics were measured in a physical property measurement system (PPMS, Quantum Design Inc.). The temperature dependence of resistance was taken with the low-frequency lock-in technique ($<10$ Hz) with a 2 $\mu$A AC current excitation. For the $V-I$ characteristics, DC voltages were measured by a Keithley 2182 nanovoltmeter and a DC current bias was applied by a Keithley 6221 current source. The critical current was extracted by first taking the derivative of $V$ v.s. $I$ data, and the current value corresponds to the first peak in $dV/dI$ v.s. $I$ was marked as the critical current. If the transitions were sharp, a constant cutoff voltage may be applied to extract the critical currents. We found that the choice of extraction methods does not lead to a significant difference in the interpretation of the data. The switching curves were employed to extract critical current, unless stated otherwise. Before zero-field measurements, the magnetic field was set to 1 T and then oscillated to zero above the transition temperature of NbSe$_2$ to minimize the effect of the remnant field on the device behavior.

\subsection*{Numerical simulations}

We simulate the current-phase relationship (CPR) of the multiferroic JJ using the following tight binding Bogoliubov-de Gennes Hamiltonian:
\begin{align}
   H^{(BdG)} & = \sum_{\mathbf{r}_n} \psi^{\dagger}_{\mathbf{r}_n} \left[ \left( (4 + q^2/4)t - \mu + U_{dip}({\bf r}_n) + U_{barrier}\delta_{x_n,0} \right) \tau_z \otimes \sigma_0  \right] \psi_{\mathbf{r}_n} \nonumber \\
   &+ \sum_{\mathbf{r}_n} \psi^{\dagger}_{\mathbf{r}_n} \left[ \frac{\Delta_Z^{(x)}({\bf r}_n)}{2} \tau_z \otimes \sigma_x + J_{exc}({\bf r}_n) \tau_0 \otimes \sigma_y - \Delta(\mathbf{r}_n)\tau_y \otimes \sigma_y \right] \psi_{\mathbf{r}_n} \nonumber \\
   &+ \sum_{\left< {\bf r}_n,{\bf r}_m \right>} \delta_{y_n,y_m} \psi^{\dagger}_{\mathbf{r}_n} \left( -t \tau_z\otimes \sigma_0 + i \alpha_R \tau_0 \otimes \sigma_x + itq_y h({\bf r}_n, {\bf r}_m) \tau_0 \otimes \sigma_z \right)
   \psi_{\mathbf{r}_m} \nonumber \\
   &+ \sum_{\left< {\bf r}_n,{\bf r}_m \right>} \delta_{x_n ,x_m} \psi^{\dagger}_{\mathbf{r}_n} \left( -t \tau_z\otimes \sigma_0 - i \alpha_R \tau_z \otimes \sigma_y + itq_x h({\bf r}_n, {\bf r}_m) \tau_0 \otimes \sigma_z \right)
   \psi_{\mathbf{r}_m},
   \label{eq:tbham}
\end{align}
where 
$
\psi_{\mathbf{r}_n} = (c_{\mathbf{r}_n \uparrow},c_{\mathbf{r}_n \downarrow},c^{\dagger}_{\mathbf{r}_n \uparrow},c^{\dagger}_{\mathbf{r}_n \downarrow})^T,
$
and $c^{\dagger}_{\mathbf{r}_n,\rho \sigma}$ ($c_{\mathbf{r}_n,\rho \sigma}$) is the creation (annihilation) operator for an electron at site $\mathbf{r}_n$ with spin $\sigma$, and $\tau_i$ and $\sigma_i$ are Pauli matrices.
The JJ is defined by the regions
$
-L_{x,s}-L_{x,n}/2 \le x \le L_{x,s} + L_{x,n}/2
$
and
$
-L_{y}/2 \le y \le L_{y}/2.
$
Then we have
\begin{align}
    U_{dip}({\bf r}) =& \begin{cases}
    \kappa_{dip} (\bf P \cdot r), & -L_{x,n}/2 \le x \le L_{x,n}/2 \\
    0, & otherwise
    \end{cases} \\
    \Delta_Z^{(x)}({\bf r}) =& \begin{cases}
    \Delta_Z^{(x)}, & -L_{x,n}/2 \le x \le L_{x,n}/2 \\
    0, & otherwise
    \end{cases} \\
    J_{exc}({\bf r}) =& \begin{cases}
    J_{exc} &  \vert x\vert \le L_{x,n}/2+\xi_{exc} \\
    0, & otherwise
    \end{cases} \\
    \Delta({\bf r}) =& \begin{cases}
    \Delta e^{i \phi/2}, & x < -L_{x,n}/2 \\
    \Delta e^{-i \phi/2}, & x > L_{x,n}/2 \\
    0, & otherwise
    \end{cases} \\
    h({\bf r}_n,{\bf r}_m) =& \begin{cases}
    1 &  \vert x_n\vert,\vert x_m\vert \le L_{x,n}/2+\xi_{exc} \\
    0, & otherwise
    \end{cases},
\end{align}
where $\xi_{exc}$ is the characteristic length scale of the exchange proximity effect in the superconducting leads. To calculate the CPR, we diagonalize the tight-binding Hamiltonian to solve for eigenvalues $\{ \epsilon_n(\phi) \}$. At temperature $T$, the CPR for a short ballistic junction is~\cite{Golubov2004}
\begin{align}
    I_{s}(\phi,T) & = - \frac{2e}{\hbar} \sum_n \tanh{\left( \frac{\epsilon_n}{2 k_B T} \right)} \frac{d\epsilon_n}{d\phi}.
\end{align}
In our simulations, we take the superconducting gap to be constant and only consider $k_B T \le 0.05 \Delta$ where the suppression of $\Delta$ according to BCS theory is negligible. 
Simulation results are presented in Extended Data Fig.~\ref{fig:abs} and \ref{fig:wideJJ}.

Here, we discuss the effect of an electric polarization $\bf P$ in a multiferroic JJ on its SDE. Numerically, we approximate the electric polarization with an effective dipole approximation resulting in an electric potential
$
U_{dip} = \kappa_{dip} (\bf P \cdot r),
$
where $\kappa_{dip}$ characterizes the permeability of the multiferroic layer.
Owing to the perfect conductivity of the superconducting electrodes, we consider an electric polarization confined to the normal region of the JJ. Furthermore, we constrain $\bf P \times q \parallel + \bf z$ as is required for spin-spiral multiferroic ordering.
Extended Data fig.~\ref{fig:ABSmultiJJ}a-c show the Andreev bound state spectra of the multiferroic JJ with RSOC.
Extended Data fig.~\ref{fig:ABSmultiJJ}d shows the $T=0$ CPR with $\vert {\bf P} \vert >0$. 
At $T=0$ and for $\pm \bf P \parallel x$, we find 
that flipping the sign of $\bf P \parallel x$ does not affect the diode rectification efficiency or polarity. Here, the asymmetry introduced in the junction by $\bf P$ leads to an asymmetric normal resistance, similar to typical ferroelectric diodes.~\cite{Blom1994}. On the other hand, for $\pm \bf P \parallel y$,
flipping $\bf P$ results in a change in the CPR since $\bf q$ along the current direction is simultaneously flipped. In general, the tunability of the CPR with $\bf P$ depends on the details of the junction and a more systematic study is needed to determine how to optimize the electric tunability of the CPR by manipulating $\bf P$.

\subsection*{Depairing momentum analysis}

%=============================================
%                 FIGURE
%=============================================

We can consider the heuristic argument given by Yuan and Fu~\cite{yuan_supercurrent_2022} to explore the diode effect in the superconducting helimagnet as it relates to finite Cooper pair momentum associated with a current bias. Consider the effect of a depairing momentum $\mathbf{\ell}$ on the energy spectrum of the superconducting helimagnet where the Hamiltonian in Eq.~(\ref{eq:ham}) is replaced by
\begin{align}
    h_{BdG}(k,\mathbf{\ell}) = \begin{pmatrix}
        h(k+\mathbf{\ell}/2) - \mu & \Delta \\
        \Delta & \mu - T^{-1} h(k - \mathbf{\ell}/2) T
    \end{pmatrix}.
\end{align}
Here we focus on $\mathbf{\ell} = \ell_x \bf x$.
The key to the heuristic argument given by Yuan and Fu for a Rahsba superconductor with an in-plane Zeeman field is that an asymmetry in the closing of the spectral gap (manifestation of the diode effect) arises when the Zeeman field is \textit{perpendicular} to $\mathbf{\ell}$ (i.e. current direction). Mathematically, this condition is a consequence of the form of the spin-orbit coupling e.g. for $\mathbf{\ell} = \ell_x \bf x$, the spin-orbit term and Zeeman terms in the Hamiltonian are aligned $\sim \left(\ell_x + \Delta_{Z}^{(y)} \right)\sigma_y$, effectively shifting the depairing momentum, see Extended Data Fig.~\ref{fig:depairing_q}a-b. Hence, if the depairing momentum term in the spin-orbit interaction is perpendicular to the Zeeman field, then there is no asymmetry in the closing of the spectral gap with $\mathbf{\ell}$. Given the general form of the spin-orbit interaction and effective Zeeman splitting in a helimagnet in the absence of an external magnetic field, we see that it is not possible to observe an asymmetry in the closing of the spectral gap with $\mathbf{\ell}$.
%, see Extended Data Fig.~\ref{fig:depairing_q}c-f. 
This implies centrosymmetric superconducting helimagnets generally will not show a diode effect associated with a depairing momentum mechanism. 
Now, if we consider Rashba spin-orbit coupling (RSOC) as discussed in the main text, we find that a non-reciprocal critical current develops in our simulations.
Incorporating RSOC into the helical superconductivity analysis above, the SDE in the depairing momentum emerges as an asymmetric suppression of the gap with the depairing momentum, see Extended Data Fig.~\ref{fig:depairing_q}c-f.

To picture the even-in-$H$ SDE observed in the experiment, it's helpful to consider how $h_Z$ affects the spectral gap of a superconducting helimagnet with RSOC. Introducing a depairing momentum into the BdG Hamiltonian will result in an indirect gap closing in the dispersion, as discussed above. Phenomenologically, the indirect gap closing in superconductors with SDE will be asymmetric in the depairing momentum~\cite{yuan_supercurrent_2022}. Now, the Zeeman effect tends to suppress the spectral gap of a superconductor with spin-singlet pairing due to a spin population imbalance. 
In our case, we find SDE will be symmetric in $\Delta_Z$ when $h_Z$ causes the spectral gap to close \textit{directly} i.e. it does not contribute to an indirect spectral gap suppression favoring a finite depairing momentum.
This is shown explicitly in Extended Data Fig.~\ref{fig:depairing_q}g-j where the effect of the depairing momentum is symmetric in $\Delta_Z^{(x)}$ whereas this ideal symmetry is lifted with $\Delta_Z^{(z)}$.

\section*{Acknowledgement}
H.Y.Y. thanks Fazel Tafti at Boston College for generously providing the access to his laboratory facilities for the growth of NiI$_2$ single crystals. H.Y.Y. thanks Margarita Davydova, Adolfo O. Fumega, Yi Tseng, Connor A. Occhialini, Jonathan Gaudet, and Ilya Sochnikov for the fruitful discussions.
J.J.C. thanks Enrico Rossi for helpful discussions and Fran\c{c}ois L\'{e}onard for his critical reading of the manuscript. K.L.W. acknowledges the support of the U.S. Army Research Office MURI program under Grants No. W911NF-20- 2-0166 and No. W911NF-16-1-0472. 
J.J.C. is supported by a LDRD. 
We acknowledge the use of the Nano and Pico Characterization Lab in the California NanoSystems Institute at UCLA.
Sandia National Laboratories is a multi-mission laboratory managed and operated by National Technology \& Engineering Solutions of Sandia, LLC (NTESS), a wholly owned subsidiary of Honeywell International Inc., for the U.S. Department of Energy’s National Nuclear Security Administration (DOE/NNSA) under contract DE-NA0003525. This written work is authored by an employee of NTESS. The employee, not NTESS, owns the right, title and interest in and to the written work and is responsible for its contents. Any subjective views or opinions that might be expressed in the written work do not necessarily represent the views of the U.S. Government. The publisher acknowledges that the U.S. Government retains a non-exclusive, paid-up, irrevocable, world-wide license to publish or reproduce the published form of this written work or allow others to do so, for U.S. Government purposes. The DOE will provide public access to results of federally sponsored research in accordance with the DOE Public Access Plan. K.W. and T.T. acknowledge support from the JSPS KAKENHI (Grant Numbers 21H05233 and 23H02052) and World Premier International Research Center Initiative (WPI), MEXT, Japan.

\section*{Author contributions}

H.Y.Y. and K.L.W. conceived the project and designed the experiments. H.Y.Y. synthesized the NiI$_2$ crystals. Y.L., S.H., and C.W.C. synthesized the NbSe$_2$ crystals. T.T. and K.W. provided and characterized bulk h-BN crystals. H.Y.Y. and C.E. performed atomic force microscopy measurements. G.Q. fabricated the bottom contact electrodes. H.Y.Y. fabricated the NiI$_2$ JJ, NbSe$_2$/NbSe$_2$, and graphite/NiI$_2$/graphite devices. A.J.B. and H.Y.Y. fabricated the Gr JJ device. H.Y.Y. performed the transport measurements. H.Y.Y. and J.J.C. analyzed the transport data. J.J.C. developed the theoretical model and performed the numerical simulations. H.Y.Y., J.J.C., and K.L.W. wrote the manuscript with inputs from all authors.

\section*{Competing interests} The authors declare no competing interests. 

\section*{Data and materials availability} All data are available are available from the corresponding authors upon request.

\begin{appendices}

\section*{Extended Data}\label{extd}
\clearpage

\begin{figure}[h!!!!]
\renewcommand\figurename{Extended Data Fig.}
    \centering
    \includegraphics[width=0.9\linewidth]{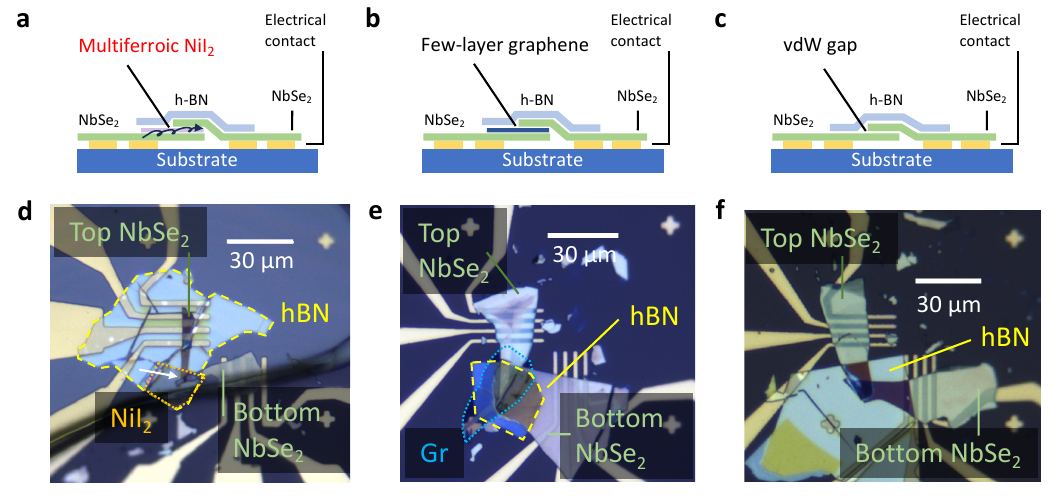}
    \caption{\textbf{Schematics and optical images of the reported superconducting devices.} \textbf{a-c}, Side views of NiI$_2$ JJ, Gr JJ, and NbSe$_2$/NbSe$_2$ devices. \textbf{d-f}, Optical images of NiI$_2$ JJ, Gr JJ, and NbSe$_2$/NbSe$_2$ devices. The junction area of NiI$_2$ JJ with a lateral dimension of 2-3 $\mu$m is pointed out by the white arrow. The thickness of the following flakes was measured by AFM: NiI$_2$ in panel d: 2.8(2) nm (4ML); few-layer graphene in panel e: 4.0(2) nm. The thickness of the following flakes is estimated from the thickness of the flakes with similar color contrast measured by AFM: top NbSe$_2$ in panel d: $\sim$20 nm; bottom NbSe$_2$ in panel d: 30-40 nm; top NbSe$_2$ flake in panel e: $\sim$20 nm; bottom flake in panel e: 60-70 nm; both NbSe$_2$ flakes in panel f: $\sim$30 nm. Multiple colors can sometimes be seen in a single flake, and the thickness above provided the best estimation of the thickness near the junction area. }
    \label{fig:extd_device}
\end{figure}
\clearpage

\begin{figure}[h!!!!]
\renewcommand\figurename{Extended Data Fig.}
    \centering
    \includegraphics[width=0.7\linewidth]{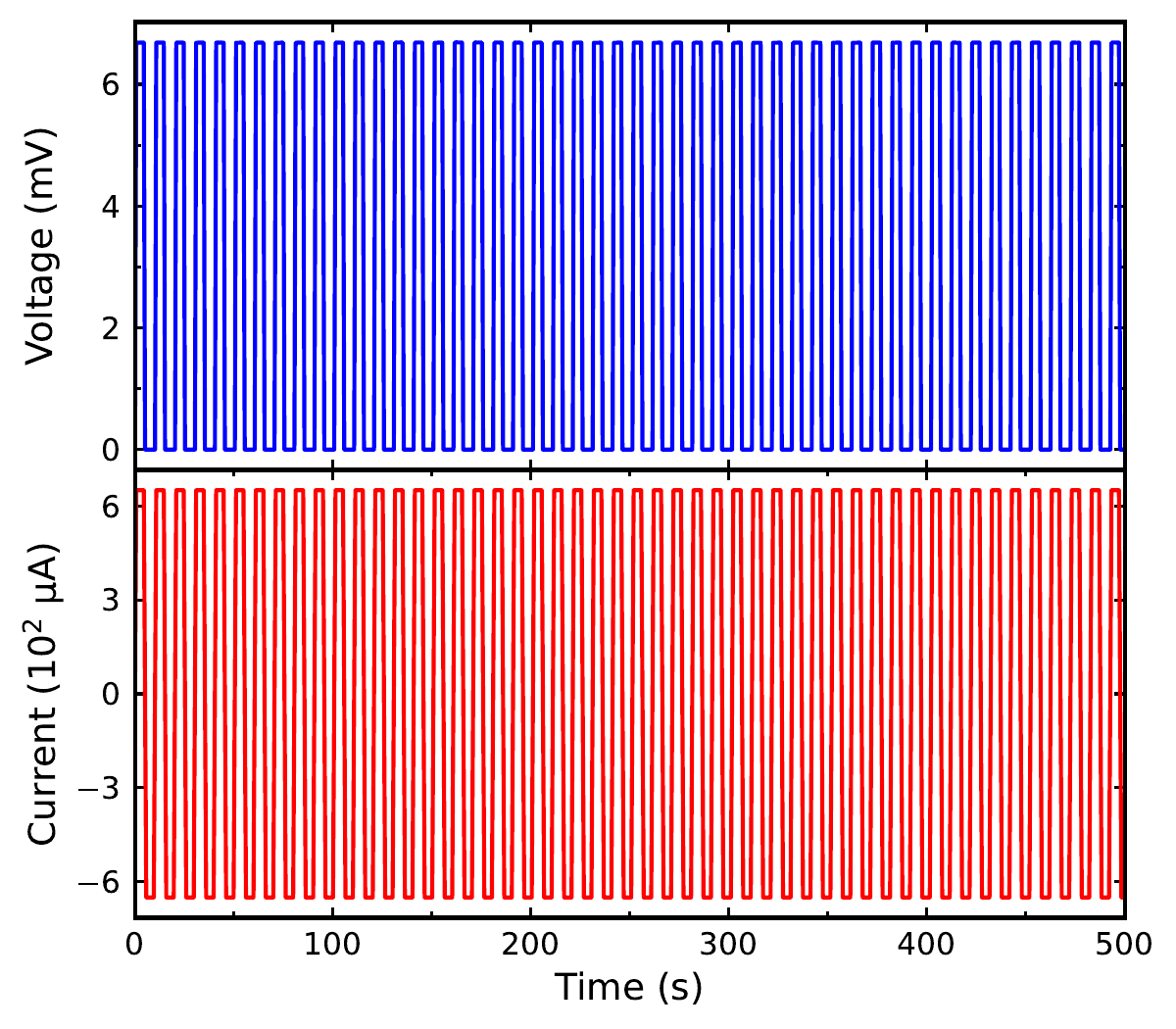}
    \caption{\textbf{Robust supercurrent switching in the NiI$_2$ device.} Demonstration of supercurrent switching with $I_\text{bias}=\pm650$ $\mu$A used in Fig. \ref{fig:fig1}d for 50 cycles. }
    \label{fig:extd_switching}
\end{figure}
\clearpage

\begin{figure}[h!!!!]
\renewcommand\figurename{Extended Data Fig.}
    \centering
    \includegraphics[width=\linewidth]{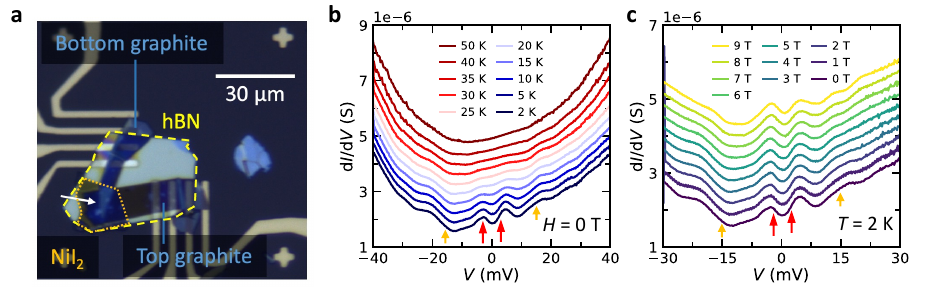}
    \caption{\textbf{Inelastic electron tunneling experiments on the graphite/NiI$_2$/graphite tunnel junction.} \textbf{a}, The optical image of the graphite/NiI$_2$/graphite tunnel junction device. The junction area is pointed out by the white arrow. The thickness of the NiI$_2$ flake at the junction is 10.5 nm as measured by AFM. The thickness of the top and bottom graphite flakes is about 20 nm and 40 nm based on the AFM results of other flakes of similar color contrast. \textbf{b}, Tunneling conductance v.s. dc voltage bias measured at different temperatures at zero field. The data taken at temperatures other than $2$ K are shifted vertically for clarity. The red arrows ($\sim \pm$ 3-5 meV) and orange arrows ($\sim \pm 15$ meV) represent a symmetric step-like increase of tunneling conductance, potentially enhanced by collective excitations \cite{klein_probing_2018}. The inferred excitation energies agree with the electromagnons and magnons energies in NiI$_2$ as detected by Raman spectroscopy and Terahertz spectroscopy \cite{song_evidence_2022,kim_terahertz_2023}. The onset temperature of collective excitations agrees with the ordering temperature of NiI$_2$ in the thin layer limit, supporting their magnetic origin \cite{song_evidence_2022}. \textbf{c}, Tunneling conductance v.s. dc voltage bias measured at different in-plane magnetic fields. The data taken at fields other than $0$ T are shifted vertically for clarity. The features representative of collective excitations in NiI$_2$ persist all the way to $H_\parallel = 9$ T, suggesting that the multiferroic order in thin-layer NiI$_2$ survives to high magnetic fields as in the bulk limit \cite{kurumaji_magnetoelectric_2013}.}
    \label{fig:extd_iets}
\end{figure}
\clearpage

\begin{figure}[h!!!!]
\renewcommand\figurename{Extended Data Fig.}
    \centering
    \includegraphics[width=\linewidth]{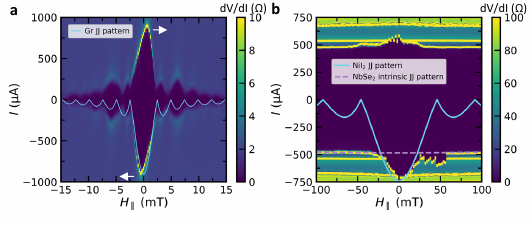}
    \caption{\textbf{Fraunhofer interference patterns of Gr JJ and NiI$_2$ JJ.} \textbf{a}, Critical current mapping of the Gr JJ at $T=2$ K. The white arrows point out the tilt of the central maxima towards opposite directions due to the cross-junction geometry (see Fig. \ref{fig:fig1}) \cite{ferrell_self-field_1963,yamashita_magneticfield_1967,stuehm_diffraction_1974}. The tilt of the interference pattern is directly linked to the anti-symmetric field dependence of SDE in Gr JJ shown in Fig. \ref{fig:fig2}. Other than the tilt, the field dependence of the critical current agrees with the Fraunhofer interference pattern as expected from a dc Josephson effect, $I_c(H)=I_c(0)\lvert \sin(\pi \Phi/\Phi_0)/(\pi \Phi/\Phi_0) \rvert = I_c(0) \lvert \text{sinc} (\Phi/\Phi_0) \rvert$, where $\Phi_0$ is the magnetic flux quantum, and $\Phi$ is the magnetic flux going through the JJ \cite{tinkham2004introduction}. Typically, $\Phi$ is calculated as $H_\parallel \times (d+2 \lambda) \times W$, where $d$ is the separation between the two superconducting electrodes, $\lambda$ is the London penetration depth of the superconductor, and $W$ is the lateral width of the junction. Here, we use $\Phi=H_\parallel \times d_\text{eff} \times W$ to model the interference patterns in our vdW JJs. With $W$ determined from the optical image of the device, the simulated curve (cyan line) showing a good agreement is generated with $d_\text{eff}=40$ nm, similar to other vdW JJs reported in the literature \cite{yabuki_supercurrent_2016,idzuchi_unconventional_2021,kang_van_2022}. We also note that if the flux focusing effect \cite{suominen_anomalous_2017} plays a role, a slightly modified expression  $I_c(H)=I_c(0)\lvert \text{sinc} (\Gamma \Phi/\Phi_0) \rvert$ with $\Gamma=3$ and $d_\text{eff}=14$ nm can also lead to a decent agreement between the observed and simulated patterns. \textbf{b}, Critical current mapping of the NiI$_2$ JJ at $T=2$ K. At low fields, a typical interference pattern ($I_c(0) \lvert \text{sinc} (\Phi/\Phi_0) \rvert$) as expected from the NiI$_2$ JJ was observed (cyan solid line, $W=3$ $\mu$m and $d_\text{eff}=15$ nm). At high fields, the pattern was truncated by the other interference pattern caused the interlayer Josephson coupling between each layer within NbSe$_2$ (purple dashed line). The coexistence of both types of interference patterns was also observed in Bi$_2$Sr$_2$CaCu$_2$O$_{8+x}$/Bi$_2$Sr$_2$CaCu$_2$O$_{8+x}$ vdW JJ \cite{zhao_time-reversal_2023}. }
    \label{fig:extd_Fpattern}
\end{figure}
\clearpage

\begin{figure}[h!!!!]
\renewcommand\figurename{Extended Data Fig.}
    \includegraphics[width=0.7\columnwidth]{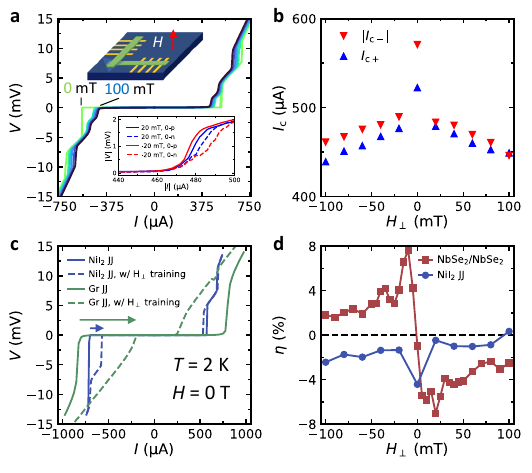}
    \centering
    \caption{\textbf{Out-of-plane magnetic field dependence of supercurrent diode effect in the NiI$_2$ JJ.} \textbf{a}, $V-I$ characteristic of the NiI$_2$ JJ with 0 mT $<H_\perp<$ 100 mT, with a 20 mT field increment at $T=2$ K. Inset: Representative switching curves at $H_\perp=\pm 20$ mT showing negative supercurrent diode effect in both cases. \textbf{b}, Critical current $I_{c+}$ and $\lvert I_{c-}\rvert$ as a function of $H_\perp$. At zero field, it is noted that the $\Delta I_c$ as deduced from panel b is smaller than the corresponding zero-field value from Fig. \ref{fig:fig2}b, due to the reduction of critical current when the measurement is taken after an out-of-plane field history. \textbf{c}, $V-I$ characteristic of NiI$_2$ JJ and Gr JJ with and without an $H_\perp$ training history. The short blue and long green arrows show the reduction of $I_{c-}$ after an $H_\perp$ training for NiI$_2$ JJ ($-100$ mT) and Gr JJ ($-10$ mT), respectively. The reduction is likely due to trapped Abrikosov vortices induced by $H_\perp$ in the NbSe$_2$ electrodes \cite{pippard_trapped_1997}, different from the Josephson vortices induced by $H_\parallel$. \textbf{d}, Comparison of $\eta$ as a function of $H_\perp$ between the NbSe$_2$/NbSe$_2$ junction and the NiI$_2$ JJ. The NbSe$_2$/NbSe$_2$ homostructure can be viewed as a superconductor with enhanced inversion symmetry breaking by stacking two NbSe$_2$ flakes. It shows an SDE with anti-symmetric out-of-plane field dependence, as expected from superconductors without inversion symmetry \cite{hou_ubiquitous_2023}. The NiI$_2$ JJ defies this general trend and shows a predominantly symmetric field dependence with a persistent negative $\eta$.}
    \label{fig:extd_oopH}
\end{figure}
\clearpage

\begin{figure}[h!!!!]
\renewcommand\figurename{Extended Data Fig.}
    \centering
    \includegraphics[width=\linewidth]{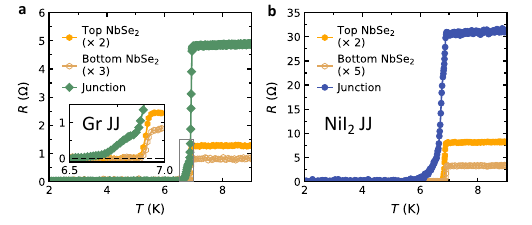}
    \caption{\textbf{Temperature dependence of the resistance across the junction and flakes.} \textbf{a}, Four-terminal resistance of the top NbSe$_2$ flake, the bottom NbSe$_2$ flake, and across the junction in the Gr JJ. The inset shows the resistance near the superconducting transition of NbSe$_2$ (marked by the gray square). \textbf{b}, Four-terminal resistance of top NbSe$_2$ flake, bottom NbSe$_2$ flake, and across the junction in the NiI$_2$ JJ. In both JJs, the resistance of both top and bottom NbSe$_2$ flakes sharply drops to zero at the transition temperature $\sim 7$ K \cite{kershaw_preparation_1967}, while the junction resistance drops to zero at a lower temperature, consistent with the typical behavior of JJs \cite{pal_josephson_2022,qiu_emergent_2023}. It is noted that the resistance of both JJs decreases monotonically as the temperature approaches the transition temperature, suggesting the high quality of our device compared to a non-homogeneous superconductor which would show an anomalous peak above the transition temperature \cite{vaglio_explanation_1993}. }
    \label{fig:extd_RT}
\end{figure}
\clearpage

\begin{figure*}[h!!!!]
\renewcommand\figurename{Extended Data Fig.}
    \centering
    \includegraphics[width=0.7\linewidth]{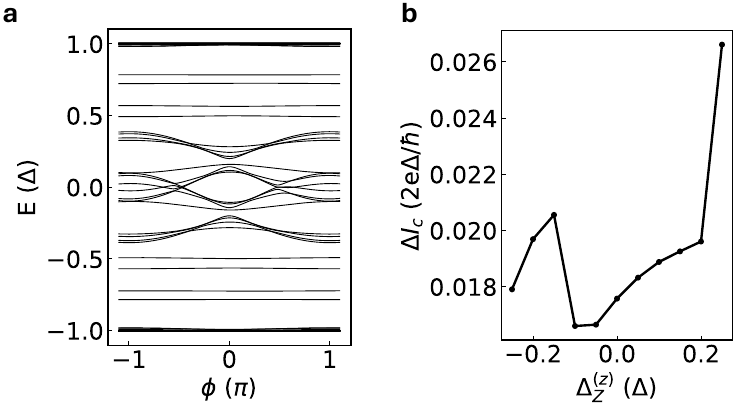} 
    \caption{\label{fig:abs}
    \textbf{Additional simulations of a helimagnet JJ.} \textbf{a}, Andreev bound state spectrum of the helimagnet JJ with CPR corresponding to Fig.~\ref{fig:theory}c with RSOC for ${\bf q} \parallel \bf x$. 
    \textbf{b}, $\Delta I_c$ as a function of $\Delta_Z^{(z)}$ (external Zeeman spin-splitting along $\bf z$ in Fig. \ref{fig:theory}a) with a mixed even-odd functional dependence. Parameters used in simulations are: $\Delta = 0.4 t$, $\mu = 1.57 t$, $\alpha_R = 0.004ta$, $J_{exc} = 0.3t$, $U_{barrier}= 4t$, $\vert \mathbf{q} \vert = 0.01 \frac{\pi}{a}$, $L_{x,s} = 300a$, $L_{x,n} = 3a$, $L_y = 10a$, and $\xi_{exc} = 5a$ where $t = \frac{\hbar^2}{2m^* a^2}$ and $a$ is the tight binding lattice constant.
    }
\end{figure*}
\clearpage

\begin{figure*}[h!!!!]
\renewcommand\figurename{Extended Data Fig.}
    \centering
    \includegraphics[width=0.95\linewidth]{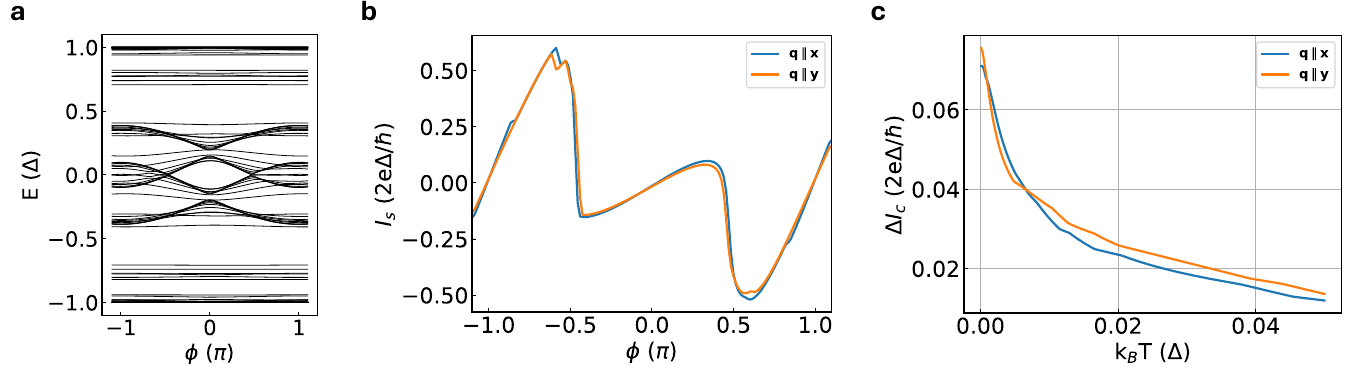} 
    \caption{\label{fig:wideJJ}
    \textbf{Simulation of a wide helimagnet JJ.}
    Simulation of a wide helimagnet JJ with $L_y = 20 a$.
    \textbf{a}, Andreev bound state spectrum for ${\bf q} \parallel \bf x$ with RSOC.
    \textbf{b}, Current-phase relationships. and
    \textbf{c}, Temperature dependence of $\Delta I_c$.
    }
\end{figure*}
\clearpage

\begin{figure*}[h!!!!]
\renewcommand\figurename{Extended Data Fig.}
    \centering
    \includegraphics[width=0.6\linewidth]{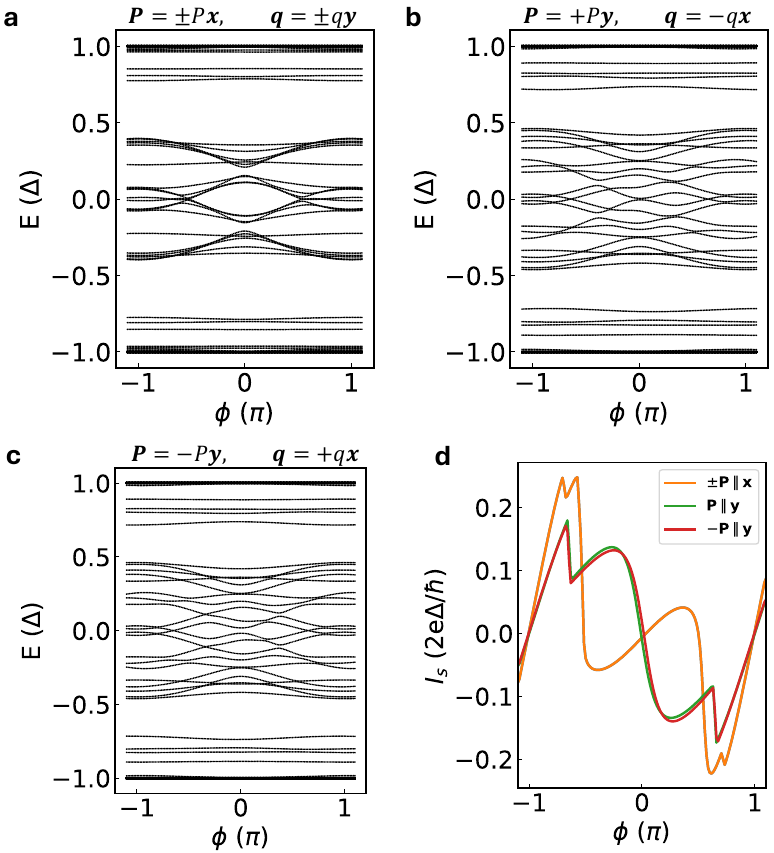} 
    \caption{\label{fig:ABSmultiJJ}
    \textbf{Simulations of multiferroic JJs.}
    \textbf{a}, Andreev bound state spectra of multiferroic JJs for ${\bf P} = \pm P \bf x$ and
    \textbf{b-c} ${\bf P} = \pm P \bf y$ with $\bf P \times q \parallel + \bf z$ (see Fig. \ref{fig:theory}a). \textbf{d}, CPRs of multiferroic JJs calculated from their corresponding spectra shown in panel \textbf{a-c}.
    }
\end{figure*}
\clearpage

\begin{figure*}[h!!!!]
\renewcommand\figurename{Extended Data Fig.}
    \centering
    \includegraphics[width=0.95\linewidth]{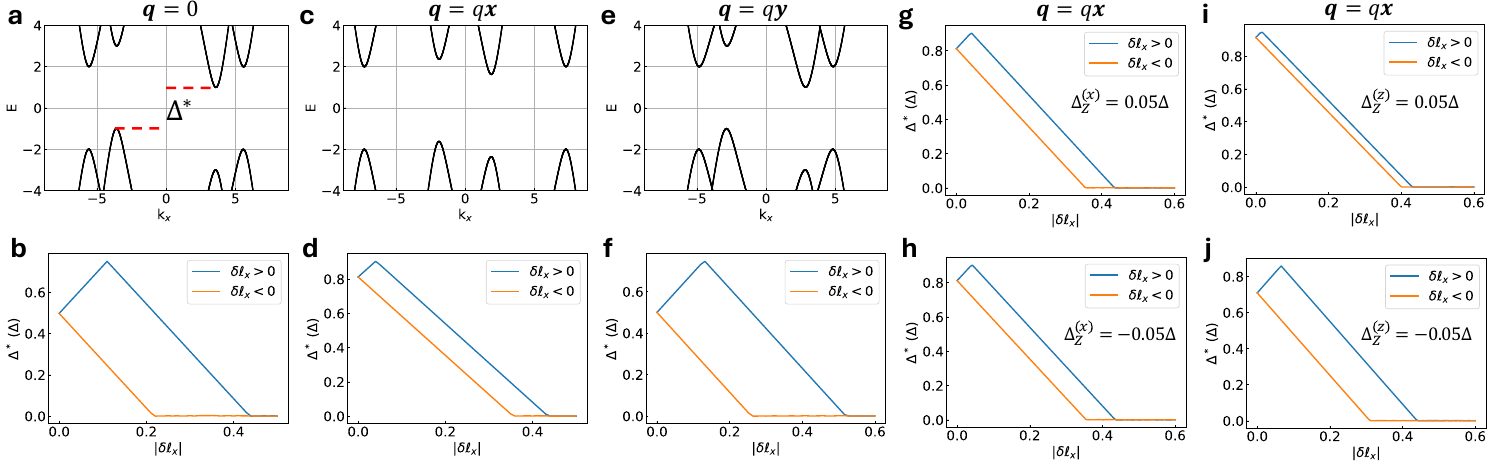} 
    \caption{\label{fig:depairing_q}
    \textbf{Single-$\bf \ell$ helical superconductivity with a helimagnet.}
    Dispersion with $\ell_x = \ell_{x,0}$ and gap dependence on depairing momentum $\delta \ell_x = \ell_{x,0}- \ell_{x}$, where $\ell_{x,0}$ is the equilibrium momentum, for a
    \textbf{a-b}, helical Rashba superconductor,
    \textbf{c-f}, helimagnet superconductor with Rashba spin-orbit coupling.
    \textbf{g-h}, Spectral gap suppression with depairing momentum for positive (top) and negative (bottom) Zeeman splitting along the x-direction. \textbf{i-j}, Spectral gap suppression with depairing momentum for positive (top) and negative (bottom) Zeeman splitting along the z-direction.
    }
\end{figure*}
\clearpage

\end{appendices}

\end{document}